\documentclass[journal=jpccck,manuscript=article]{achemso}
\usepackage[version=3]{mhchem}
\usepackage{graphicx}
\usepackage{bm}
\usepackage{natbib}

\def\eV{\,\textrm{eV}}
\def\V{\,\textrm{V}}

\title{Oxygen reduction activity on perovskite oxide surfaces:\\
a comparative first-principle study of LaMnO$_3$, LaFeO$_3$ and LaCrO$_3$}

\author{Yan Wang}
\author{Hai-Ping Cheng}
\affiliation{Department of Physics and Quantum Theory Project, University of Florida, Gainesville,
Florida 32611, USA} \email{hping@ufl.edu}

\begin{document}

\begin{abstract}
The understanding of oxygen reduction reaction (ORR) activity on perovskite oxide surfaces is essential
for promising future fuel cell applications. We report a comparative study of ORR mechanisms on
La$B$O$_3$ ($B$=Mn, Fe, Cr) surfaces by first-principles calculations based on density functional
theory (DFT). Results obtained from varied DFT methods such as generalized gradient approximation
(GGA), GGA+$U$ and the hybrid Hartree-Fock density functional method are reported for comparative
purposes. We find that the results calculated from hybrid-functional method suggest that the order
of ORR activity is LaMnO$_3$ $>$ LaCrO$_3$ $>$ LaFeO$_3$, which is in better agreement with recent
experimental results (Suntivich \textit{et al.}, Nature Chemistry 3, 546 (2011))
than those using the GGA or GGA+$U$ method.
\end{abstract}

\maketitle \pagebreak

\section{\label{sec:1}INTRODUCTION}
Oxygen reduction reaction (ORR), one of electrochemical energy conversion processes, plays a very
important role in renewable energy technologies, including fuel cells and metal-air batteries.
Searching for a highly active catalyst to replace noble metal cathodes is driven by the need of
efficient and low-cost ORR catalysts, which are essential for mass marketing a fuel cell
technology to address the world's energy needs.

Recent experiments by Suntivich \textit{et al.} \cite{Suntivich:2010, Suntivich:2011} shows that
perovskite transition-metal oxides can exhibit high electrocatalytic activity for ORR in alkaline
electrolytes. It is also suggested that ORR activity on perovskite oxide surfaces is related to the
$e_g$ occupation in the $B$-site cation, which is indicative of the strength of bonding between
transition metal ion and adsorbed oxygen. A moderate amount of $e_g$-filling in perovskite oxides
such as LaMnO$_3$ ($e_g=1$) yields higher activity as compared to other oxides with either too
little (LaCrO$_3$ with $e_g=0$) or too much $e_g$ electron filling (LaFeO$_3$ with $e_g=2$).

ORR on La$B$O$_3$ perovskite oxide surfaces has been theoretically investigated extensively in the
past using first-principles methods based on density functional theory
\cite{Choi:2007,Kotomin:2008,Lee:2009,Choi:2009,Kushima:2010,Chen:2011}. All of these studies
considered atomic and molecular oxygen adsorption on the La$B$O$_3$ surface only for its application in
solid oxide fuel cells. However, in alkaline fuel cells the ORR involves a more complicated
reaction pathway, in which ORR intermediates such as hydroxides and peroxides \cite{Suntivich:2011}
will form on the surface. Furthermore, in most studies the standard (semi) local DFT methods, local
density approximation or generalized gradient approximation (GGA), are used, while only a few of
them use GGA+$U$ approach to describe strongly correlated electrons\cite{Lee:2009,Lee:2011}. The
role of GGA+$U$ is to address on-site Coulomb interactions in the localized orbitals (such as
$d$ orbitals in transition metals) with an additional Hubbard-type term $U$
\cite{Anisimov:1991,Liechtenstein:1995}. It has been found that La$B$O$_3$ surface energetics
are strongly depend on the parameter $U$ used in the calculation \cite{Lee:2009}.


These recent studies further motivate us to use higher-level theoretical treatments, the so-called
``parameter-free'' hybrid functional approach, for a complete description of the ORR on the
La$B$O$_3$ perovskite oxide surfaces.
A hybrid functional, a combination of exact nonlocal
orbital-dependent Hartree-Fock exchange and a standard local exchange-correlation functional,
provides a significant improvement over the GGA description and enables accurate computation of
electronic properties and energetics of molecular systems as well as extended systems including
transition metal oxides \cite{Paier:2005,Batista:2006,Marsman:2008} without the need for
system-dependent adjustable parameters or decisions of which electrons to localize. So far ORR
activity on perovskite oxide surfaces is unexplored by hybrid-functional approaches. In the present
work we aim at filling this gap.

In this paper we report a comparative study of ORR activity on La$B$O$_3$ ($B$=Mn, Fe, Cr)
perovskite oxide surfaces using first-principles calculations. Calculations based on various DFT
functionals such as GGA, GGA+$U$ and hybrid functional are performed for comparative purposes. From
binding energies of ORR intermediates, we obtain free energy changes at each step of the ORR and
present overall free energy diagrams for each surface. The kinetics of ORR on each perovskite
oxide surface are found to differ significantly with different DFT methods used in the calculation.
We find that the hybrid functional method yields better agreement with recent experiments by
Suntivich \textit{et al.} \cite{Suntivich:2010, Suntivich:2011} while the results from the GGA and
GGA+$U$ methods fail to explain the experimental observations.

\section{\label{sec:2}Computational Methods}

\subsection{\label{sec:2.1}DFT calculations}

Our calculations are performed using the plane-wave-basis-set Vienna \textit{ab initio} simulation
package VASP\cite{VASP:1996} (version 5.2). The projector-augmented-wave (PAW) methods are used
to describe the interactions between atomic cores and valence electrons, with a kinetic energy cutoff
of $500 \eV$ employed in all simulations. The $3p$-semi-core states are treated as valence states for
transition metal atoms Cr ($3p^6 3d^5 4s^1$), Mn ($3p^6 3d^5 4s^2$), and Fe ($3p^6 3d^6 4s^2$). For
the exchange-correlation functional we used the GGA method with Perdew-Burke-Ernzerhof (PBE)
formulation \cite{Perdew:1996}. We also apply the GGA+$U$ method to reduce the self-interaction
error and improve the description of correlation effects. For the GGA+$U$ calculation we make use
of the standard Dudarev implementation \cite{Dudarev:1998} where the on-site Coulomb interaction
for the localized orbitals is parametrized by $U_{\rm eff}=U-J$ using the PBE functional. We apply
the optimized effective interaction parameter $U_{\rm eff}$ for the metal atoms in La$B$O$_3$
($U_{\rm eff} = 4$, 4 and $3.5 \eV$ for Mn, Fe and Cr, respectively), determined by fitting the
enthalpies of the oxidation reactions \cite{Wang:2006}. These values have previously been shown to
provide a description of La$B$O$_3$ electronic structure that is in good agreement with the available
experimental data \cite{Lee:2009}. Furthermore, we perform single-point total energy calculations
with the hybrid functional approach developed by Heyd-Scuseria-Ernzerhof (HSE06) \cite{Heyd:2006} for
the approximation of the exchange-correlation energy and potential. In the HSE06 approach, one
quarter of the PBE short-range exchange is replaced by the exact Hartree-Fock exchange, and the full
PBE correlation energy is included. The range-separation parameter is set to be $0.2 \, \textrm{\AA}^{-1}$.

\begin{figure}
{\includegraphics[width=10cm]{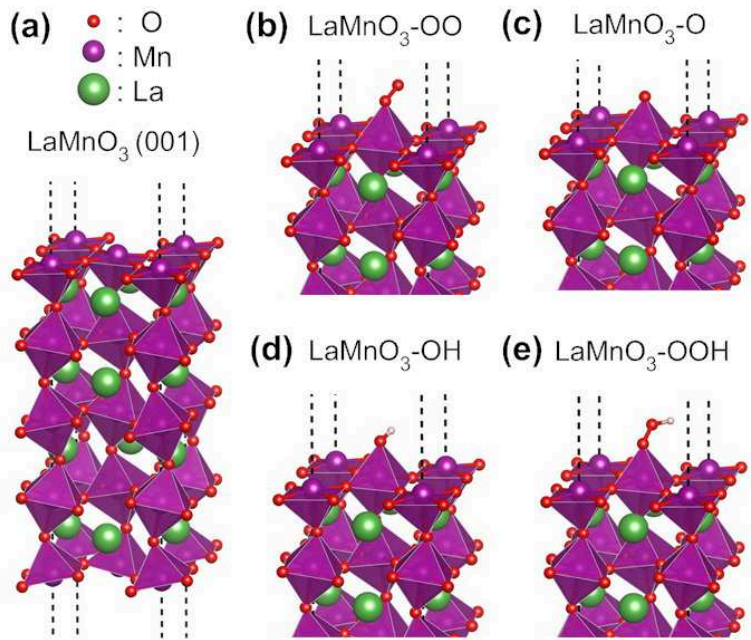}} \caption{\label{fig:geometry}(Color online).
Optimized geometry of bare LaMnO$_3$ (001) surface (a) and surfaces with adsorbed ORR intermediates
OO* (b), O* (c), HO* (d), and HOO* (e) intermediates. The dashed lines are the boundaries of the
supercell.}
\end{figure}

The La$B$O$_3$ perovskite oxide surfaces are simulated with a slab including nine atomic layers
with transition metal $B$-terminated (001) surfaces at each side. Such surfaces are chosen simply
because the (001) surfaces are generally the most stable in perovskites and the redox active
transition metal atoms are expected to be responsible for ORR catalytic activity
\cite{Suntivich:2011}. The distorted perovskite structure of \textit{pnma} is considered to be the
structure identified in the experiment of Ref.~\citet{Suntivich:2011}, and the equilibrium bulk
geometry with optimized lattice constants is used to construct the La$B$O$_3$ slab geometries. A
large vacuum spacing of $ 15 \, \textrm{\AA} $ was used between two slabs to prevent interaction
between the system with its images in adjacent unit cells. We use the $1 \times 1$ surface unit
cell, and two-dimensional periodic boundary conditions are applied for the surface directions with
a $6 \times 6 $  Monkhorst-Pack type of $k$-point sampling \cite{Monkhorst:1976}. The four bottom
atomic layers of the slab are fixed at the optimized bulk lattice constant, while the top five
layers as well as the ORR intermediate on the surface are fully relaxed. In all calculations a
dipole correction is applied. The ORR intermediates are modeled by the same slab with the
intermediate adsorbed at the $B$ sites on the top surface of the slab. Adsorbates placed above the
surfaces correspond to a coverage of 50\%. The geometries are optimized until the force on each
atom falls below the convergence criterion of $ 0.02 \eV$/{\AA}. The optimized structures of bare
La$B$O$_3$ surfaces and surfaces with adsorbed ORR intermediates OO*, O*, HO* and HOO* (asterisk
denoting adsorbed species hereafter), taking LaMnO$_3$ as an example, are shown in
Fig.~\ref{fig:geometry} (a)-(e). For the HSE06 calculations, four bottom atomic layers of each
La$B$O$_3$ slab are removed in order to save computational effort.

The magnetic structures of La$B$O$_3$ perovskites are complicated due to phase transitions at
different temperatures. For simplicity and consistency we apply the ferromagnetic ordering for all
the studied La$B$O$_3$ perovskites, by an initial assignment of non-zero (parallel) spin
to all the $B$ atoms. The self-consistent procedure determining the electron and spin density
distribution for each geometry are always carried out leaving the total spin unconstrained and free
to evolve. We expect our assumption will not introduce large errors in our energetic results, as we
focus on the adsorption energy and relative free energy of the ORR intermediates on the surface, in
which the total energy difference caused by changing the relative spin orientations inside the slab
should have negligible effect.


\subsection{\label{sec:2.2}Free energy diagrams for oxygen reduction reaction}

The following four-electron ORR reaction pathway for La$B$O$_3$ at the cathode in alkaline
electrolytes has been suggested \cite{Suntivich:2011} (overall process $\rm O_2 + 2H_2O + 4e^-
\rightarrow 4 OH^-$):\\ \\
1) surface hydroxide displacement as:
\begin{equation}
\textrm{La}{B} \textrm{O}_3\textrm{-OH}+ \textrm{O}_2+e^-
 \longrightarrow  \textrm{La}B\textrm{O}_3\textrm{-OO}+\textrm{OH}^- \\
\end{equation}
2) surface peroxide formation as:
\begin{equation}
\textrm{La}B\textrm{O}_3\textrm{-OO}+\textrm{H}_2\textrm{O}+e^-
 \longrightarrow \textrm{LaBO}_3\textrm{-OOH}+ \textrm{OH}^- \\
\end{equation}
3) surface oxide formation as:
\begin{equation}
\textrm{La}B\textrm{O}_3\textrm{-OOH}+e^-
 \longrightarrow \textrm{La}B\textrm{O}_3\textrm{-O}+\textrm{OH}^- \\
\end{equation}
and 4) surface hydroxide regeneration as:
\begin{equation}
\textrm{La}B\textrm{O}_3\textrm{-O}+\textrm{H}_2\textrm{O}+e^-
 \longrightarrow  \textrm{La}B\textrm{O}_3\textrm{-OH}+\textrm{OH}^-
\end{equation}

The free energy change of each ORR reaction step is calculated based on a computational hydrogen
electrode model suggested by N{\o}rskov \textit{et al}. \cite{Norskov:2004}. This method has been
shown to predict trends for the oxygen reduction reaction on metals quite well \cite{Norskov:2004}.
In this method the potential affects the relative free energy through the chemical potential of the
electrons in the electrode. We convert the calculated DFT energies into Gibbs free energies by
adding entropic ($TS$) and zero-point energy (ZPE) corrections to the ORR intermediates, so that
\begin{equation}
\Delta G=\Delta E+\Delta \textrm{ZPE}-T\Delta S+\Delta G_\Phi
\end{equation}
where $\Delta E$ is the calculated DFT reaction energy, $\Delta \textrm{ZPE}$ is the change in ZPE
and $\Delta S$ is the change in the entropy. ZPE corrections and entropies of the ORR intermediates
are calculated from the vibrational frequencies according to standard methods, and those of the
gas-phase molecules are obtained from thermodynamics databases. $\Delta G_\Phi$ is the effect of
electrode potential which is applied by shifting the free energy change $\Delta G$ by $\Delta
G_\Phi=e\Phi$, where $e$ is the elementary charge and $\Phi$ is the potential difference between
electrode and counter electrode (versus reversible hydrogen electrode, RHE). The equilibrium
potential $\Phi_{\rm eq}$ corresponds to zero net reaction free energy ($\sum_{i=1}^4 G_i = 2\Delta
G_{\rm W}+4e\Phi=0$)  of the overall ORR process, thus we have $\Phi_{\rm eq}=-\Delta G_{\rm
W}/(2e)$, where $\Delta G_{\rm W}=G({\rm H_2O})-G({\rm H_2})-G({\rm O_2})/2$ is the free energy of
formation of water from $\rm H_2$ and $\rm \frac12 O_2$. The ORR potential $\Phi_{\rm ORR}$
corresponds to the highest potential $\Phi$ at which all steps along the reaction decrease the free
energy. The theoretical ORR overpotential is then calculated by $\eta=\Phi_{\rm eq}-\Phi_{\rm
ORR}$.


Furthermore, effect of water on the ORR activity is also taken into account by a solvation
correction to ORR intermediates, since reactions occur in the presence of water in electrolytes
\cite{Suntivich:2011}. The solvation corrections are obtained from a previous study of
platinum-catalyzed ORR using the Poisson-Boltzmann implicit continuum model \cite{Sha:2010}. The
solvation correction energies are $-0.32$, $-0.47$, $-0.75$ and $-0.54 \eV $
for OO*, O*, HO*, and HOO* intermediates, respectively.


\section{\label{sec:3}Results and discussion}

\subsection{\label{sec:3.1}Surface Energetics and Electronic properties}

\begin{table*}
\caption{\label{tab:adenergy} Calculated DFT binding energy of each ORR intermediate on La$B$O$_3$
surfaces (unit: eV). The DFT binding energies of the ORR intermediates are calculated according to
$\Delta E_{\rm adsorbate}=E_{\rm LaBO_3\textrm{-}adsorbate}-E_{\rm LaBO_3}-E_{\rm adsorbate}$. The
absorbtion energies $\Delta E^{\rm ad}_{\rm adsorbate}$ relative to energies of H$_2$O and H$_2$
are also given (in parentheses, see text for definition).}
\begin{tabular}{cccccccc}
structures & method & $\Delta E_{\rm OH}$ ($\Delta E^{\rm ad}_{\rm OH}$) & $\Delta E_{\rm OO}$ &
  $\Delta E_{\rm OOH}$ ($\Delta E^{\rm ad}_{\rm OOH}$) & $\Delta E_{\rm O}$ ($\Delta E^{\rm ad}_{\rm O}$)\\
\hline  & GGA & $-2.927$ (0.417) & $-0.659$ & $-1.424$ (3.739) & $-3.755$ (1.913)\\
LaMnO$_3$ & GGA+$U$ & $-2.107$ (1.236) & $-0.205$ & $-0.724$ (4.439) & $-1.987$ (3.681)\\
 & HSE06 & $-1.647$ (1.355) & 0.162 & $-0.482$ (4.844) & $-1.278$ (4.015)\\
\hline  & GGA & $-2.686$ (0.658) & $-0.491$ & $-1.224$ (3.939) & $-3.457$ (2.211)\\
LaFeO$_3$ & GGA+$U$ & $-2.287$ (1.057) & $-0.199$ & $-0.794$ (4.369) & $-2.134$ (3.534)\\
 & HSE06 & $-1.778$ (1.223) & 0.328 & $-0.028$ (5.298) & $-1.368$ (3.926)\\
 \hline  & GGA & $-3.276$ (0.068) & $-0.931$ & $-1.786$ (3.377) & $-4.986$ (0.682)\\
LaCrO$_3$ & GGA+$U$ & $-2.429$ (0.915) & $-0.172$ & $-1.014$ (4.149) & $-3.515$ (2.153)\\
 & HSE06 & $-1.858$ (1.144) & ~0.305 & $-0.678$ (4.648) & $-2.972$ (2.322)\\
\end{tabular}
\end{table*}

We first calculate the DFT binding energy for each ORR intermediate on the La$B$O$_3$ (001)
surfaces. The DFT binding energies of the ORR intermediates are calculated according to $\Delta
E_{\rm adsorbate}=E_{\rm LaBO_3\textrm{-}adsorbate}-E_{\rm LaBO_3}-E_{\rm adsorbate}$, where
$E_{\rm LaBO_3\textrm{-}adsorbate}$, $E_{\rm LaBO_3}$ and $E_{\rm \rm adsorbate}$ are the energies
of the surface with adsorbed intermediate, the bare surface and the isolated ORR intermediate,
respectively. The negative sign of $E_{\rm adsorbate}$ corresponds to energy gain of the system
due to adsorption of ORR intermediate, and the more negative the $E_{\rm adsorbate}$ value the
stronger is the chemical interaction between the adsorbate and the surface. It should be also noted
that our calculations give the reference energies of $-9.86$ and $-17.04 \eV $ for O$_2$ with the GGA and
the hybrid functional, respectively, and no correction for this has been made in the values of this
work. While such a correction would alter the absolute binding energies it has no effect on the
relative difference in binding energies between surfaces. The results are summarized in Table
\ref{tab:adenergy} for comparison among different DFT functionals. It is clear that for each ORR
intermediate the binding energy (absolute values) decreases from the GGA value to that of GGA+$U$
and then especially strongly to the HSE06 hybrid-functional result. This holds for all three
perovskite surfaces.

Among the four ORR intermediates, the O* binding energy has the strongest dependence of the
calculating method. Taking LaMnO$_3$ as an example, the GGA binding energy $E_{\rm O}$ is about $
1.8 \eV $ larger (in its absolute value) than that of GGA+$U$, and it is about $ 2.5 \eV $ larger
than the HSE06 value. For the binding energy $E_{\rm OO}$ a relatively weaker dependence can be
found. For LaMnO$_3$ the difference between calculated values of $E_{\rm OO}$ from the GGA and
GGA+$U$ methods is about $ 0.45 \eV $, and that between the GGA and HSE06 methods is about $ 0.8
\eV $. Similar behaviors can be also found for LaFeO$_3$ and LaCrO$_3$ in Table \ref{tab:adenergy}.
These can be explained by the fact that the interaction between atomic oxygen and the surface
transition metal $B$ ion is much stronger than the other ORR intermediates, thus the effect of
correlation corrections in the the GGA+$U$ and HSE06 calculations are much more obvious on the O*
binding energies. The calculated charge transfer results further support this point. In Table
\ref{tab:charge} we show the effective Bader charges (based on the real-space-charge density
\cite{Henkelman:2006}) of transition metal ion and adsorbed ORR intermediate on each La$B$O$_3$
surface. It is clear that the charge transfer between the surface and the adsorbed ORR intermediate
is large for La$B$O$_3$-O but much smaller for La$B$O$_3$-OO. Our results are consistent with a
recent \textit{ab initio} GGA+$U$ study by Lee \textit{et al.} \cite{Lee:2009}, which also shows
that the O* binding on the La$B$O$_3$ surface has a stronger $U_{\rm eff}$ dependence than the OO*
binding. In our calculations, the even larger correction of binding energies in the HSE06 hybrid
functional results compared to the corrections in the results from the GGA+$U$ method is primarily
due to the fact that the $U_{\rm eff}$ is applied only at the transition metal $B$ atoms in the
GGA+$U$ calculations.

\begin{table*}
\caption{\label{tab:charge} Effective Bader charges $q$ (in $e$) of transition metal ions and
adsorbed ORR intermediates (in parentheses) on La$B$O$_3$ surfaces.}
\begin{tabular}{cccccccccc}
perovskites & method & bare surface & La$B$O$_3\textrm{-}\rm OO$ & La$B$O$_3\textrm{-}\rm OOH$ & La$B$O$_3\textrm{-}\rm O$ & La$B$O$_3\textrm{-}\rm OH$\\
\hline  & GGA & 1.66 & 1.78 (-0.31) & 1.81 (-0.36) & 1.80 (-0.51) & 1.82 (-0.39) \\
LaMnO$_3$ & GGA+$U$ & 1.71 & 1.76 (-0.16) & 1.81 (-0.30) & 1.86 (-0.52) & 1.89 (-0.44) \\
 & HSE06 & 1.83 & 1.91 (-0.25) & 1.96 (-0.37) & 1.93 (-0.46) & 1.99 (-0.42) \\
\hline  & GGA & 1.57 & 1.64 (-0.24) & 1.68 (-0.33) & 1.70 (-0.52) & 1.74 (-0.47)  \\
LaFeO$_3$ & GGA+$U$ & 1.79 & 1.84 (-0.15) & 1.85 (-0.35) & 1.73 (-0.54) & 1.83 (-0.47)  \\
 & HSE06 & 1.81 & 1.94 (-0.18) & 1.93 (-0.41) & 1.87 (-0.57) & 1.95 (-0.55)  \\
\hline  & GGA & 1.78 & 1.92 (-0.40) & 1.91 (-0.39) & 2.01 (-0.55) & 1.93 (-0.42) \\
LaCrO$_3$ & GGA+$U$ & 1.81 & 1.97 (-0.18) & 1.93 (-0.29) & 2.02 (-0.51) & 1.99 (-0.39) \\
 & HSE06 & 1.94 & 2.02 (-0.28) & 1.91 (-0.39) & 2.17 (-0.56) & 2.14 (-0.43) \\
\end{tabular}
\end{table*}

In comparing the calculated binding energies of the three La$B$O$_3$ systems, we notice that for
each ORR intermediate the binding energy (in its absolute value) is generally larger for LaCrO$_3$
than that of LaMnO$_3$ or LaFeO$_3$, with the only exception the OO* intermediate with the GGA+$U$
and HSE06 results. In particular, the binding energy $E_{\rm O}$ for LaCrO$_3$ is (depending on the
method) about 1.2$\sim$1.7 and 1.4$\sim$1.6 $\eV$ more negative than that of LaMnO$_3$ and
LaFeO$_3$, respectively, indicating a stronger interaction between the LaCrO$_3$ surface and the
atomic oxygen adsorbate. For the OO* intermediate, we find that though the GGA results still show a
$\sim 0.3 \eV $ larger binding energy (in its absolute value) for LaCrO$_3$ as compared to the
LaMnO$_3$ ($\sim 0.4 \eV $ larger as compared to LaFeO$_3$), the GGA+$U$ method yields very similar
binding energies of molecular oxygen for all three perovskite surfaces. The HSE06 method, however,
gives the largest binding energy $E_{\rm OO}$ with a positive sign for LaFeO$_3$ surface,
indicating a relatively weak binding ability of LaFeO$_3$ surface for OO* intermediate. This is
also true for the HOO* intermediate where the HSE06 method yields smaller binding energy $E_{\rm
OOH}$ (in its absolute value) for LaFeO$_3$ than the other two perovskites.

In Table \ref{tab:adenergy} we also show the calculated adsorption energies of HO*, HOO* and O*
intermediates relative to H$_2$O and H$_2$, according to the following reactions as defined in
Ref.~\citet{Man:2011}:
\begin{eqnarray} \label{ad}
\Delta E^{\rm ad}_{\rm OH} &=& E_{\rm LaBO_3\textrm{-}OH}-E_{\rm LaBO_3}-(E_{\rm H_2O}-
 {\textstyle\frac{1}{2}} E_{\rm H_2}) \\
\Delta E^{\rm ad}_{\rm OOH} &=& E_{\rm LaBO_3\textrm{-}OOH}-E_{\rm LaBO_3}-(2E_{\rm H_2O}-
 {\textstyle\frac{3}{2}} E_{\rm H_2}) \\
\Delta E^{\rm ad}_{\rm O} &=& E_{\rm LaBO_3\textrm{-}O}-E_{\rm LaBO_3}-(E_{\rm H_2O}-E_{\rm H_2}).
\end{eqnarray}
where $E_{\rm H_2O}$, and $E_{\rm H_2}$ are the calculated energies of H$_2$O and H$_2$ molecules
in the gas phase, respectively. Our calculated adsorption energies of HO*, HOO* and O*
intermediates with the GGA-PBE method are very close to the recent reported results by Man
\textit{et al.} \cite{Man:2011} using the GGA-RPBE method, with less than $ 0.1 \eV $ difference
for LaFeO$_3$ and LaCrO$_3$. For LaMnO$_3$ our GGA adsorption energies are slightly larger than the
results reported in Ref.~\citet{Man:2011}, with 0.55, 0.15 and $ 0.27 \eV $ difference for the O*,
HO* and HOO* intermediates, respectively. The discrepancy may be because of different
pseudopotentials (ultrasoft) used in the Ref.~\citet{Man:2011}.


\subsection{\label{sec:3.3} ORR on La$B$O$_3$ surface}

\begin{table}
\caption{\label{tab:fenergy} Calculated free energy change $\Delta G$ for each ORR step in
La$B$O$_3$ (unit: eV).
* denotes the potential-determining step for the ORR with the smallest $|\Delta G|$.}
\begin{tabular}{cccccccc}
structures & method & $\Delta G_{1}$ & $\Delta G_{2}$ & $\Delta G_{3}$ & $\Delta G_{4}$\\
\hline  & GGA & -0.642* & -0.649* & -2.370 & -1.000\\
LaMnO$_3$ & GGA+$U$ & -1.008 & -0.403* & -1.302 & -1.948\\
 & HSE06 & -0.758 & -0.526* & -1.372 & -2.165\\
\hline  & GGA & -0.715 & -0.618* & -2.272 & -1.057\\
LaFeO$_3$ & GGA+$U$ & -0.822 & -0.479* & -1.379 & -1.981\\
 & HSE06 & -0.462 & -0.237* & -1.916 & -2.206\\
 \hline  & GGA & -0.564 & -0.739 & -3.239 & -0.119*\\
LaCrO$_3$ & GGA+$U$ & -0.652* & -0.726 & -2.540 & -0.743\\
 & HSE06 & -0.405* & -0.864 & -2.870 & -0.682\\
\end{tabular}
\end{table}

The calculated free energy change of each ORR step at zero electrode potential
($\Phi=0\, \textrm{V} $ vs. RHE)
are listed in Table \ref{tab:fenergy}. $\Delta G_1$, $\Delta G_2$, $\Delta G_3$, and $\Delta G_4$
correspond to the free energy changes in the ORR reaction steps of Eqs.~(1), (2), (3) and (4),
respectively. All reaction steps are exothermic for all three perovskite oxide surfaces with
different calculation methods. However, for each surface GGA+$U$ and HSE06 methods give quite
different relations between different reaction steps respect to the GGA values.
The size of the ORR potential-determining step can be estimated according to
\begin{equation}
G={\rm min} (|\Delta G_{1}|, |\Delta G_{2}|, |\Delta G_{3}|, |\Delta G_{4}|) .
\end{equation}
In the following we further analyze the catalytic performance by calculating the
free energy diagrams under equilibrium and ORR potentials for each La$B$O$_3$ surface.

\begin{figure}
{\includegraphics[width=10cm]{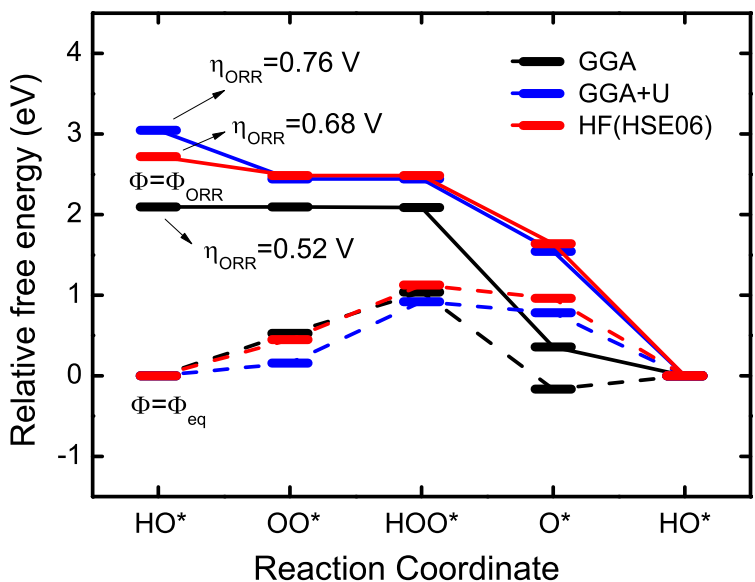}} \caption{\label{fig:Mn}(Color online). Free energy
diagrams of the ORR on the LaMnO$_3$ (001) surface. Solid and dashed lines represent reactions at
the ORR potential ($\Phi=\Phi_{\rm ORR}$) and the equilibrium potential ($\Phi=\Phi_{\rm eq}$),
respectively. The ORR potential corresponds to the highest potential $\Phi$ at which all steps
along the reaction decrease the free energy. The theoretical overpotentials $\eta$ are marked for
results with different functionals.}
\end{figure}

Fig.~\ref{fig:Mn} shows the free energy diagrams of the ORR on the LaMnO$_3$ (001) surface with
different calculation methods for representative potentials $\Phi_{\rm eq}$ and $\Phi_{\rm ORR}$.
At the equilibrium potential $\Phi_{\rm eq}$, for the GGA method it is clear that only reaction
step (3), surface O* formation, are energetically downhill (exothermic) and all other three steps
are uphill (endothermic). The O* intermediate is stabilized on the LaMnO$_3$ surface. This is very
similar to the case of ORR on Pt(111) surfaces studied by N{\o}rskov et al.
\cite{Norskov:2004,Hansen:2008}. The first two steps of the ORR process, the surface OO*/HO*
exchange and the HOO* formation, are the two potential-determining steps. The fact that these two
steps become exothermic at nearly the same ORR potential indicates that LaMnO$_3$ has a reactivity
close to optimal, as the free energy changes in the four reaction steps are linearly related; thus
increasing one of the step will decrease another. The theoretical ORR overpotential is $ 0.52 \V $,
which is very close to the theoretical overpotential of $ 0.48 \V $ for ORR on Pt(111)
\cite{Hansen:2008}. In contrast to the strongly bound O* intermediate with the GGA functional,
applying the GGA+$U$ and HSE06 functionals change the results by giving much higher relative free
energy of O* intermediate in the ORR process. This should mainly be attributed to the largely
decreased interaction strength between the O* intermediate and the LaMnO$_3$ surface for the
GGA+$U$ and HSE06 functionals with respect to the GGA result, consistent with the calculated
binding energy as shown in Table~\ref{tab:adenergy}. The HO* intermediate becomes the most stable
one on the LaMnO$_3$ surface and the surface hydroxide regeneration (step 4) becomes exothermic.
The potential limiting steps lie in the second electron transfer step of the ORR (OOH* formation),
and the theoretical ORR overpotentials are $ 0.76 \V $ and $ 0.68 \V $ for the GGA+$U$ and HSE06
results, respectively.

\begin{figure}
{\includegraphics[width=10cm]{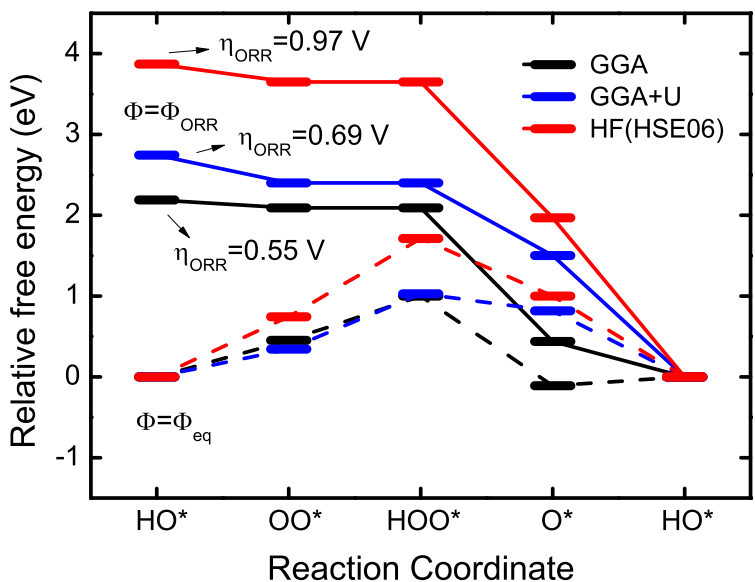}} \caption{\label{fig:Fe}(Color online). Free energy
diagrams of the ORR on the LaFeO$_3$ (001) surface.}
\end{figure}

The calculated free energy diagrams for ORR on the LaFeO$_3$ (001) surface under $\Phi_{\rm eq}$
and $\Phi_{\rm ORR}$ are shown in Fig.~\ref{fig:Fe}.
It has previously been suggested that for LaFeO$_3$ the surface OO*/HO* exchange
does not gain sufficient energy, and thus the ORR kinetics are limited by
the rate of OO*/HO* exchange (step 1) \cite{Suntivich:2011}.
However, our calculation indicates
that this reaction step is only the secondary potential-determining step.
The primary potential-determining step is located at the HOO* formation (step 2),
which has a slightly larger free energy change than that of the OO*/HO*
exchange step under equilibrium potential independent of the choice of the calculation method.
The GGA calculation for LaFeO$_3$ yields results very similar to the
calculated free energy diagram for LaMnO$_3$,
for which we obtain a low ORR overpotential of $ 0.55 \V $.
Also similar to LaMnO$_3$, the self-interaction corrections in the GGA+$U$ and HSE06
calculations lower the binding energy of O* on the LaFeO$_3$ surface and give clearly higher
relative free energy for the O* intermediate than the HO* and OO* intermediates in the ORR process.
Moreover, for the HSE06 method the free energy change in the two endothermic steps, OO*/HO* exchange
and HOO* formation, are increased as compared to the results from GGA and GGA+$U$.
As a result, while the GGA+$U$ method gives a theoretical ORR overpotential of $ 0.69 \V $,
comparable to the GGA method, the overpotential calculated for the HSE06 hybrid-functional is
$ 0.97 \V $, which is almost twice as large as that of the GGA value.

\begin{figure}
{\includegraphics[width=10cm]{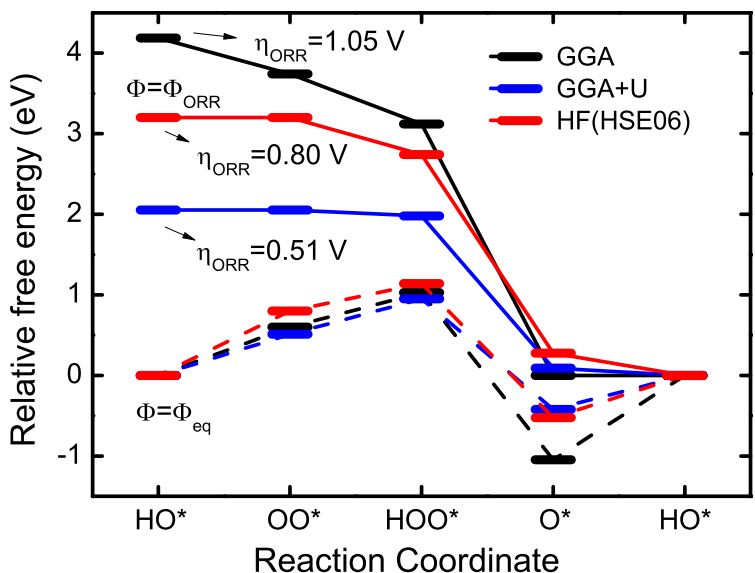}} \caption{\label{fig:Cr}(Color online). Free energy
diagrams of the ORR on the LaCrO$_3$ (001) surface.}
\end{figure}

For ORR on a LaCrO$_3$ (001) surface, at the equilibrium potential the O* intermediate is strongly
stabilized on the surface for the GGA functional, as shown in Fig. \ref{fig:Cr}. This situation
does not change much even if we apply yhe GGA+$U$ method or HSE06 functional in the calculation.
For the GGA functional it is clearly shown that the surface HO* regeneration reaction (step 4) is
the potential-determining step. The corresponding ORR overpotential is $ 1.05 \V $, much higher
than that of LaMnO$_3$ or LaFeO$_3$. This agrees very well with the experimental observations in
Ref.~\citet{Suntivich:2011}, which suggests that the O* intermediate on the LaCrO$_3$ surface is
not sufficiently destabilized and the ORR kinetics are limited by the rate of surface OH*
regeneration. However, using both the GGA+$U$ method and HSE06 functional we find the relative free
energy of O* intermediate becomes about $ 0.6 \eV $ higher with respect to the GGA result, which
makes the HO* regeneration step no longer the most endothermic step in the ORR. Instead, the
HO*/OO* exchange step (step 1) becomes the potential-determining step. For the HSE06 functional the
calculated ORR overpotential is $ 0.80 \V $, which is $ 0.12 \V $ larger than that of LaMnO$_3$ but
$ 0.17 \V $ smaller than that of LaFeO$_3$. The overpotential of $ 0.51 \V $ calculated from the
GGA+$U$ method is much lower than that of LaMnO$_3$ ($ 0.76 \V $) or LaFeO$_3$ ($ 0.69 \V $), in
conflict with the experimental observations in Ref.~\citet{Suntivich:2011}. We therefore conclude
that the GGA+$U$ method with $U_{\rm eff}$ applied on the Cr $d$ state of LaCrO$_3$ may not be a
physically appropriate approach to calculate the surface reaction energies.

\section{\label{sec:4}Summary}

In summary, we conduct a comparative first-principles study of the ORR activity of the three perovskite
oxides LaMnO$_3$, LaFeO$_3$ and LaCrO$_3$.
In addition to the extensively used GGA and GGA+$U$ methodologies,
we also apply the advanced hybrid-functional method.
We find that the calculated surface
binding energies of ORR intermediates are strongly dependent on the method, and the free energy
diagrams of ORR are described quite differently in GGA, GGA+$U$ and hybrid functional approaches,
especially for LaFeO$_3$ and LaCrO$_3$ surfaces.

We show that Cr-sites on the LaCrO$_3$ surface are better adsorption centers for atomic oxygen than
the other two transition metals, but the strong binding of the O* intermediate does not favor the
surface hydroxide (OH*) regeneration step of the ORR process. For the LaFeO$_3$ surface the
interactions between the Fe-site and the OO* and HOO* intermediates are relatively weak, and the
calculated ORR free energy diagram indicates that the HOO* formation and OO*/HO* displacement
reactions are the primary and secondary potential-determining steps, respectively. These findings
agree well with much of the $e_{g}$-filling model deduced from experiments. The results calculated
from the hybrid-functional method suggest that the order of ORR activity is $ \textrm{LaMnO}_3 >
\textrm{LaCrO}_3 > \textrm{LaFeO}_3$. This is in better agreement with recent experimental
observation in Ref. \citet{Suntivich:2011} than those from the GGA or GGA+$U$ method. The GGA
results yields similar free energy diagrams and ORR activities for LaMnO$_3$ and LaFeO$_3$, while
the GGA+$U$ suggest that LaCrO$_3$ has the lowest ORR overpotential. In neither case does the
obtained order of ORR activity agree with the experiment.

\acknowledgement This work was supported by the NSF under Grant No.~DMR-0804407. The authors
acknowledge DOE/NERSC and UF-HPC centers for providing computational resources. Figures with
geometry are produced by VESTA graphical program \cite{Vesta3}.


\begin{thebibliography}:


\bibitem{Suntivich:2010} J. Suntivich, H. A. Gasteiger, N. Yabuuchi, Y. Shao-horn, J. Electrochem. Soc. \textbf{157}, B1263 (2010)

\bibitem{Suntivich:2011} J. Suntivich, H. A. Gasteiger, N. Yabuuchi, H. Nakanishi, J. B. Goodenough, Y. S. Horn, Nature Chemistry
\textbf{3}, 546-550 (2011).


\bibitem{Choi:2007} Y. Choi, D. S. Mebane, M. C. Lin, and M. Liu, Chem. Mater. \textbf{19} 1690-1699, (2007).

\bibitem{Kotomin:2008} E. A. Kotomin, Y. A. Mastrikov, E. Heifetsa, and J. Maiera, Phys. Chem. Chem. Phys. \textbf{10}, 4644-4649
(2008).

\bibitem{Choi:2009} Y. Choi, M. E. Lynch, M. C. Lin, and M. Liu, J. Phys. Chem. C \textbf{113} 7290-7297, (2009).

\bibitem{Kushima:2010} A. Kushima, S. Yip, and B. Yildiz, Phys. Rev. B \textbf{82}, 115435 (2010).


\bibitem{Chen:2011} H.-T. Chen, P. Raghunath, and M. C. Lin, Langmuir \textbf{27}, 6787-6793 (2011).

\bibitem{Lee:2009} Y.-L. Lee, J. Kleis, J. Rossmeisl, and D. Morgan, Phys. Rev. B \textbf{80}, 224101 (2009).

\bibitem{Lee:2011} Y.-L. Lee, J. Kleis, J. Rossmeisl, Y. Shao-Horn and D. Morgan, Energy Environ. Sci, \textbf{4}, 3966 (2009).


\bibitem{Anisimov:1991} V. I. Anisimov, J. Zaanen, and O. K. Andersen, Phys. Rev. B \textbf{44}, 943 (1991).

\bibitem{Liechtenstein:1995} A. I. Liechtenstein, V. I. Anisimov, and J. Zaanen, Phys. Rev. B \textbf{52}, R5467 (1995).


\bibitem{Paier:2005} J. Paier, R. Hirschl, M. Marsman, and G. Kresse, J. Chem. Phys. \textbf{122}, 234102 (2005).

\bibitem{Batista:2006} E. R. Batista, J. Heyd, R. G. Hennig, B. P. Uberuaga, R. L. Martin, G. E. Scuseria, C. J. Umrigar, and J. W.
Wilkins, Phys. Rev. B \textbf{74}, 121102(R) (2006).

\bibitem{Marsman:2008} M. Marsman, J. Paier, A. Stroppa and G. Kresse, J. Phys. Condens. Matter \textbf{20}, 064201
(2008).

\bibitem{VASP:1996} G. Kresse and J. Furthm\"{u}ller, Comput. Mat. Sci. \textbf{6}, 15 (1996).

\bibitem{Perdew:1996} J. P. Perdew, K. Burke, and M. Ernzerhof, Phys. Rev. Lett. \textbf{77}, 3865 (1996).

\bibitem{Dudarev:1998} S. L. Dudarev, G. A. Botton, S. Y. Savrasov, C. J. Humphreys, and A. P. Sutton, Phys. Rev. B \textbf{57}, 1505 (1998).

\bibitem{Wang:2006} L. Wang, T. Maxisch, and G. Ceder, Phys. Rev. B \textbf{73}, 195107 (2006).

\bibitem{Heyd:2006} J. Heyd, G. E. Scuseria, and M. Ernzerhof, J. Chem. Phys. \textbf{124}, 219906 (2006).

\bibitem{Monkhorst:1976} H. J. Monkhorst and J. D. Pack, Phys. Rev. B \textbf{13}, 5188 (1976).

\bibitem{Norskov:2004} J. K. N{\o}rskov, J. Rossmeisl, A. Logadottir, L. Lindqvist, J.R. Kitchin, T. Bligaard, and H. J\'{o}nsson, J. Phys. Chem. B 108, 17886
(2004).

\bibitem{Sha:2010} Y. Sha, T. H. Yu, Y. Liu, B. V. Merinov and W. A. Goddard III, J. Phys. Chem. Lett., \textbf{1},
856-861(2010).

\bibitem{Henkelman:2006} G. Henkelman, A. Arnaldsson, and H. J\'{o}nsson, Comput. Mater. Sci. \textbf{36},
254-360 (2006).

\bibitem{Man:2011} I. C. Man, H.-Y. Su, F. Calle-Vallejo, H. A. Hansen, J. I. Mart\'{i}nez, N. G. Inoglu, J. Kitchin, T. F. Jaramillo, J. K. N{\o}rskov, and J. Rossmeisl,
ChemCatChem \textbf{3}, 1159 (2011).

\bibitem{Hansen:2008} H. A. Hansen, J. Rossmeisl and J. K. N{\o}rskov, Phys. Chem. Chem. Phys. \textbf{10}, 3722-3730 (2008).


\bibitem{Vesta3} K. Momma and F. Izumi, J. Appl. Crystallogr. \textbf{44}, 1272 (2011).

\end{thebibliography}
\end{document}